\documentclass[preprint,onecolumn,nofootinbib]{revtex4}
\usepackage[colorlinks=true,linkcolor=blue,urlcolor=blue,filecolor=black,citecolor=red,pdfstartview=FitV,pdftitle={},pdfsubject={},pdfkeywords={},pdfpagemode=None,bookmarksopen=true]{hyperref}
\usepackage{graphicx}
\usepackage{amsmath}
\usepackage{amsfonts}
\usepackage{amssymb,ulem}
\usepackage{color,xcolor}%
\usepackage{CJK}
\usepackage{subfigure}
\usepackage{amsthm,amsmath,amssymb}
\usepackage{mathrsfs}
\usepackage{multirow}
\usepackage{dcolumn}
\usepackage{float}
\begin{document}
\title{Quasinormal overtones and grey-body factors of Frolov black holes}

\author{Ying Huang $^{1}$}
\thanks{huangyingyaya@yeah.net}
\author{Huajie Gong $^{2}$}
\thanks{huajiegong@hunnu.edu.cn, corresponding author}
\author{Li-Gang Zhu $^{1}$}
\thanks{zlgoupao@163.com}
\author{Jian-Pin Wu $^{1}$}
\thanks{jianpinwu@yzu.edu.cn, corresponding author}

\affiliation{
	$^{1}$ \mbox{Center for Gravitation and Cosmology, College of Physical Science and Technology,} \mbox{Yangzhou University, Yangzhou 225009, China}\\
	$^{2}$ \mbox{Department of Physics, Institute of Interdisciplinary Studies, Key Laboratory of} \mbox{Low Dimensional Quantum Structures and Quantum Control of Ministry of Education,} \mbox{Synergetic Innovation Center for Quantum Effects and Applications, and Hunan Research} \mbox{Center of the Basic Discipline for Quantum Effects and Quantum Technologies,} \mbox{Hunan Normal University, Changsha 410081, China}
}

\begin{abstract}

We investigate scalar quasinormal modes (QNMs), with particular attention to overtones, and grey-body factors of Frolov black holes, which interpolate between the Reissner--Nordstr\"om (RN) and Hayward limits. Over the allowed ranges of charge $q$ and regularization parameter $\alpha_0$, all QNM frequencies have negative imaginary parts, confirming stability against scalar perturbations. Increasing either $q$ or $\alpha_0$ can trigger overtone outbursts, with the Frolov threshold shifted to smaller parameter values than in the RN or Hayward cases. Angular momentum does not eliminate the outburst but postpones its onset from $n=1$ at $l=0$ to $n\geq 2$ at $l=1$. Comparisons of the effective potential show that, for $l=1$, charge raises the barrier peak much more strongly than $\alpha_0$, while the clearest deviations from the Hayward and RN  references are localized near the horizon. Using the QNM--grey-body correspondence at $l=1$, we find that overtone corrections are essential, and the amended formula substantially outperforms the eikonal truncation. At fixed frequency, both $q$ and $\alpha_0$ suppress transmission, with charge being the dominant factor, consistent with the stronger barrier enhancement under the charge scan.

\end{abstract}

\maketitle

\section{Introduction}\label{sec-intro}

Classical general relativity (GR) predicts spacetime singularities in the core of black holes (BHs), where curvature invariants diverge and the theory breaks down in the ultraviolet regime \cite{Penrose:1964wq,Joshi:2011rlc}. This breakdown motivates extensive efforts to construct regular (non-singular) black hole (RBH) spacetimes that replace the singular core with a finite-curvature interior while preserving the external BH geometry \cite{dymnikova1992vacuum,Ayon-Beato:1998hmi,Bronnikov:2000vy,Ansoldi:2008jw}.

Based on their core structure, RBH solutions are commonly divided into two classes. De Sitter-core RBHs exhibit positive Riemann curvature near the center. The Bardeen metric \cite{2767662} pioneered this approach, with notable subsequent models including the Hayward \cite{Hayward:2005gi} and Frolov \cite{Frolov:2016pav} spacetimes. Minkowski-core RBHs, by contrast, feature asymptotically vanishing curvature at the center \cite{Xiang:2013sza,Culetu:2013fsa,Rodrigues:2015ayd,Simpson:2019mud,Ghosh:2014pba}. Comprehensive reviews of these models and their phenomenological implications are available in \cite{Bambi:2023try,Vagnozzi:2022moj,Lan:2023cvz}.

Among de Sitter-core RBHs, the Frolov BH offers a particularly useful arena for studying the interplay between charge and quantum-gravity-inspired regularization. Proposed by Frolov in 2016 \cite{Frolov:2016pav}, this spacetime introduces a length-scale parameter $\alpha_0$---interpreted as the Hubble length and tied to an effective cosmological constant $\Lambda=3/\alpha_0^2$---that regularizes curvature divergences while preserving asymptotic Schwarzschild or Reissner-Nordstr\"om (RN) behavior. The Frolov BH is a charged generalization of the Hayward spacetime \cite{Hayward:2005gi}. It reduces to the Hayward BH as $q\to 0$ and to the RN BH as $\alpha_0\to 0$. Unlike either limiting case alone, the Frolov model simultaneously accommodates charge and a universal regularization hair, making it well suited for comparing the competing effects of electromagnetic and near-core modifications. In the near-core regime ($r\to 0$), the geometry approaches a de Sitter-like core with curvature invariants bounded by $\sim\alpha_0^{-2}$, consistent with Markov's limiting curvature conjecture \cite{Markov:1984ii}, while horizon structure and many thermodynamic properties of classical BHs are retained \cite{Ahmed:2025cfv}. Studies of the Frolov BH have explored shadows, lensing, and thermodynamics \cite{Kumar:2019pjp,Gohain:2024piy,Gohain:2024aod,Bouzenada:2025ypa,Kala:2025xnb,Song:2024kkx}. However, dynamical probes based on quasinormal modes (QNMs) have so far focused mainly on the fundamental spectrum \cite{Lo:2025njp,Yang:2025dbn,Xia:2023zlf,Quan:2026awx,Zhu:2024wic,Tan:2024qij,Guo:2023nkd,Liu:2022dcn}. If deviations from classical predictions are imprinted on ringdown signals, higher overtone modes may prove more sensitive than the fundamental mode alone.

This overtone sensitivity is reinforced by recent gravitational-wave observations by LIGO/Virgo, which have underscored the role of higher overtone modes in refining estimates of remnant BH parameters and enhancing the signal-to-noise ratio of ringdown signals \cite{London:2014cma,LIGOScientific:2016lio,Baibhav:2017jhs,Giesler:2019uxc,Giesler:2024hcr}. Overtones are especially sensitive to geometry near the event horizon, making them valuable probes of deviations from classical predictions \cite{Konoplya:2022iyn,Konoplya:2022hll,Konoplya:2022pbc,Konoplya:2023kem,Konoplya:2023hqb,Hirano:2024fgp}. In regular BH backgrounds, higher overtone modes can exhibit an ``outburst'' phenomenon, characterized by a sudden drop in the real part of the frequency beyond a critical value of the regularization or charge parameter \cite{Konoplya:2022pbc,Konoplya:2023kem,Gong:2023ghh,Zhang:2024nny}. This phenomenon reflects subtle near-horizon modifications that are largely invisible to the fundamental mode.

In \cite{Song:2024kkx}, we analyzed the fundamental QNMs of the Frolov RBH under scalar perturbations. That study showed that fundamental modes always lie in the lower half-plane, indicating stability under scalar perturbations, and that quantum-gravity effects slow the decay of these modes. For $l=0$, the real part of the fundamental frequency varies non-monotonically with $\alpha_0$, whereas for $l\geq 1$ this non-monotonicity is suppressed, suggesting that angular momentum dominates over the quantum-gravity correction at the fundamental level. The overtone spectrum of the charged Frolov BH, however, remains largely unexplored. In related work on the Hayward BH, Gong \textit{et al.} \cite{Gong:2023ghh} identified overtone outbursts driven by quantum corrections, but a systematic study of how both charge $q$ and $\alpha_0$ shape the overtone hierarchy in the Frolov spacetime is still missing. Key open questions include how overtone outbursts compare quantitatively with those in the RN and Hayward limits, whether angular momentum $l$ suppresses outbursts in lower overtones and forces one to examine higher $n$, and how near-horizon changes in the effective potential trigger these transitions.

Complementarily, a correspondence between QNMs and grey-body factors has recently been established \cite{Konoplya:2024lir,Konoplya:2024vuj}. Grey-body factors govern the absorption cross section of Hawking radiation and wave transmission across the potential barrier surrounding a BH. In the eikonal limit, the transmission coefficient $\Gamma_l$ can be approximated using only the fundamental QNM frequency $\omega_0$, but for small angular momentum the first overtone $\omega_1$ must be included to achieve adequate accuracy. This correspondence has been tested for fields with spin $s=0,1,2$ in several spacetimes \cite{Konoplya:2024lir,Konoplya:2024vuj}, yet its application to regular BHs with competing charge and quantum-gravity parameters requires reliable overtone data that are presently unavailable for the Frolov case.

The remainder of this paper is organized as follows. Section~\ref{secModel} presents the Frolov spacetime, scalar-field perturbations, and the pseudospectral method for computing QNFs. Section~\ref{secOvertone} reports overtone results and analyzes their significance. Section~\ref{secGBQNMs} discusses grey-body factors and their connection with QNMs. Conclusions and further discussions are given in Section~\ref{secconclusions}.

\section{Model and numerical method}\label{secModel}

This section establishes the complete framework for computing scalar QNMs of the Frolov BH: the spacetime and parameter conventions, the perturbation equation and QNM definition, and the pseudospectral method used in the following sections.

\subsection{Frolov geometry}\label{secModelMetric}

The static, spherically symmetric Frolov metric \cite{Frolov:2016pav} reads
\begin{eqnarray}
	&&
	ds^2=-f(r)dt^2+\frac{1}{f(r)}dr^2+r^2 d\theta^2+r^2 \sin^2 \theta d\phi^2\,,
	\label{Frolov-metric-v1}
	\\
	&&
	f(r)=1-\frac{(2Mr-q^2)r^2}{r^4+(2Mr+q^2)\alpha_0^2}\,,
	\label{Frolov-metric-v2}
\end{eqnarray}
where $M$ and $q$ denote the BH mass and charge, respectively. The regular core is characterized by an effective cosmological constant $\Lambda=3/\alpha_0^2$, with $\alpha_0$ interpreted as the Hubble length and bounded by \cite{Hayward:2005gi}
\begin{equation}
	\alpha_0\leq \sqrt{16/27}M\,.
	\label{Hayward-constrain}
\end{equation}
As $\alpha_0 \to 0$, the Frolov BH reduces to the RN BH, and as $q \to 0$, it reduces to the Hayward BH \cite{Hayward:2005gi}.  

For $r\to\infty$, the metric function approaches
\begin{equation}
	f(r)=1-\frac{2M}{r}+\frac{q^2}{r^2}+O(r^{-4})\,,
	\label{Frolov-asymptotic}
\end{equation}
so that $q$ carries the same meaning as the electric charge in the RN limit and the spacetime is asymptotically flat. At the origin the regular core satisfies $f(0)=1$, connecting a de~Sitter-like interior to the exterior RN geometry.

Event horizons are located at the positive roots of $f(r)=0$, equivalently
\begin{equation}
	r^4-2Mr^3+q^2r^2+2M\alpha_0^2 r+q^2\alpha_0^2=0\,,
	\label{Frolov-horizon}
\end{equation}
which yields an inner horizon $r_{h-}$ and an outer horizon $r_{h+}$ with $r_{h-}<r_{h+}$ in the sub-extremal regime. In the RN limit ($\alpha_0\to 0$) the extremal configuration occurs at $q=M$, where $r_{h-}=r_{h+}$. In the Hayward limit ($q\to 0$) extremality is reached at $\alpha_0=\sqrt{16/27}\,M$, i.e., at the upper bound of Eq.~\eqref{Hayward-constrain} \cite{Hayward:2005gi}. When both $q$ and $\alpha_0$ are nonzero, increasing either parameter drives the spacetime toward extremality earlier than in the corresponding RN or Hayward limit, narrowing the parameter window that admits two distinct horizons.

\subsection{Test scalar field and effective potential}\label{secModelPert}

A test scalar field $\Phi$ on this background satisfies the Klein-Gordon equation
\begin{equation}
	\frac{1}{\sqrt{-g}}\partial_\nu(g^{\mu\nu}\sqrt{-g}\partial_\mu\Phi)=0\,,
	\label{scalar_eq}
\end{equation}
where $g$ is the metric determinant and $g^{\mu\nu}$ the inverse metric. Spherical symmetry allows the separation
\begin{equation}\label{separate}
	\Phi(t,r,\theta,\phi)=\sum_{l,m}\frac{\Psi_{l,m}(t,r)}{r}Y_{l,m}(\theta,\phi)\,,
\end{equation}
with spherical harmonics $Y_{l,m}$ and angular quantum numbers $l$ and $m$. For fixed $l$ and $m$ we write $\Psi_{l,m}\equiv\Psi$. Eq.~\eqref{scalar_eq} then takes the Schr\"odinger-like form
\begin{equation}\label{Sch_like_eq}
	-\frac{\partial^2\Psi}{\partial t^2}+\frac{\partial^2\Psi}{\partial r_{\ast}^2}-V_{\text{eff}}\Psi = 0\,,
\end{equation}
with
\begin{equation}\label{V_eff1}
	V_{\text{eff}}=f(r)\left(\frac{l(l+1)}{r^2}+\frac{f'(r)}{r}\right)\,,
\end{equation}
and tortoise coordinate $r_*$ defined by $dr_{*}/dr=1/f(r)$.

\subsection{Pseudospectral discretization}\label{secModelPS}

We employ the pseudospectral method \cite{Jansen:2017oag,Wu:2018vlj,Fu:2018yqx,Gong:2023ghh}, which has been used successfully for overtone spectra in regular BH backgrounds. We use units in which $M=1$. Using the static time coordinate $t$, we set $\Psi=e^{-i\omega t}\psi$, so that Eq.~\eqref{Sch_like_eq} becomes
\begin{equation}\label{KGeqution}
\frac{\partial^2\psi}{\partial r_*^2}+(\omega^2-V_{\text{eff}})\psi=0\,.
\end{equation}
For asymptotically flat BHs the QNM boundary conditions are
\begin{equation}
	\psi \sim e^{\pm i \omega r_*}\,, \quad r_*\rightarrow \pm \infty\,,
	\label{BC-infinity}
\end{equation}
ingoing at the horizon and outgoing at infinity. We then introduce the ingoing Eddington--Finkelstein coordinate $v=t+r_*$ and $u=1/r$, and factor out the boundary asymptotic behavior. Denoting the redefined radial function again by $\psi(u)$, we obtain
\begin{eqnarray}
		&&
		(-(l^2+l)u^2-2iu(1-(2+2u)f(u)+(u+2u^2)f'(u))\omega+(4+8u)(1-f(u)
		-2uf(u))\omega^2)\psi(u)
		\nonumber
		\\
		&&
		+(u^4f'(u)+2iu^2(1-(2+4u)f(u))\omega)\psi'(u)+u^4 f(u) \psi''(u)=0\,,
		\label{waveEQ-v2}
\end{eqnarray}
with
\begin{equation}
	f(u)=1+\frac{u(-2+q^2 u)}{1+2u^3 \alpha_{0}^2+q^2 u^4 \alpha_{0}^2}\,.
	\label{fv2}
\end{equation}

Here $f(u)$ denotes $f(r=1/u)$, and primes in Eq.~\eqref{waveEQ-v2} denote derivatives with respect to $u$. For numerical discretization, we rescale $u\to u/u_h$, where $u_h=1/r_{h+}$, so that spatial infinity and the event horizon correspond to $u=0$ and $u=1$, respectively.
The rescaled equation is discretized using the Chebyshev grid on $[0,1]$ \cite{Jansen:2017oag}:
\begin{equation}
	\label{Cg-Lcf}
	\begin{aligned}
	&u_i=\frac{1+\cos(i\pi/N)}{2}\,,\quad i=0,\ldots,N\,,\\
	&C_j(u)=\prod_{i=0,i\neq j}^N \frac{u-u_i}{u_j-u_i}\,.
	\end{aligned}
\end{equation}
Then, the field $\psi(u)$ can be approximated as
\begin{equation}
	\psi(u)\approx\sum_{j=0}^N \psi(u_j)C_j(u)\mathrm,
\end{equation}
where the cardinal functions are linear combinations of Chebyshev polynomials $T_n(2u-1)$ of the first kind:
\begin{equation}
	\begin{gathered}
	C_j(u)=\frac2{Np_j}\sum_{m=0}^N\frac1{p_m}T_m(2u_j-1)T_m(2u-1)\,,\\
	p_0=p_N=2\,,\quad p_j=1\ (1\leq j\leq N-1)\,.
	\end{gathered}
\end{equation}

Following the discretization procedure of Ref.~\cite{Jansen:2017oag}, we evaluate the coefficient functions on the grid points and replace derivatives with respect to the rescaled coordinate $u$ by the corresponding derivative matrices. Since Eq.~\eqref{waveEQ-v2} is quadratic in $\omega$, the resulting matrix equation takes the form
\begin{equation}\label{eq:qnm-quadratic}
	(M_0+\omega M_1+\omega^2 M_2)\boldsymbol{\psi}=0\,,
\end{equation}
where $\boldsymbol{\psi}=(\psi(u_0),\ldots,\psi(u_N))^{\mathsf T}$ and $M_k$ ($k=0,1,2$) are purely numerical matrices obtained by collecting the terms of order $\omega^k$ in the discretized equation.
Following the companion linearization of Ref.~\cite{Jansen:2017oag}, we introduce an auxiliary vector $\boldsymbol{\chi}$ and write the equivalent generalized eigenvalue problem as
\begin{equation}\label{eq1}
	\begin{pmatrix}
		M_0 & M_1\\
		0 & I
	\end{pmatrix}
	\begin{pmatrix}\boldsymbol{\psi}\\ \boldsymbol{\chi}\end{pmatrix}
	=
	\omega
	\begin{pmatrix}
		0 & -M_2\\
		I & 0
	\end{pmatrix}
	\begin{pmatrix}\boldsymbol{\psi}\\ \boldsymbol{\chi}\end{pmatrix},
\end{equation}
with $I$ the $(N+1)\times(N+1)$ identity matrix. The lower block row enforces $\boldsymbol{\chi}=\omega\boldsymbol{\psi}$, and substituting this relation into the upper row recovers Eq.~\eqref{eq:qnm-quadratic}. The two block matrices have dimension $2(N+1)\times 2(N+1)$. We obtain the QNFs by solving Eq.~\eqref{eq1} and retaining the finite eigenvalues that converge under changes in grid resolution.

For each $l$, the modes are ordered by increasing $|\omega_{\mathrm{Im}}|$, and the overtone index $n=0,1,2,\ldots$ is assigned accordingly.

\section{Analysis of overtone modes}\label{secOvertone}

We now present numerical overtone spectra obtained with the method of Sec.~\ref{secModel}. With $M=1$, we first vary charge $q$ while comparing Frolov BHs with RN ($\alpha_0=0$) and Hayward ($q=0$) limits, then vary $\alpha_0$. Throughout the scans, all imaginary parts remain negative, confirming stability under scalar perturbations.

\begin{figure}[H]
	\centering{
		\includegraphics[height=4.6cm]{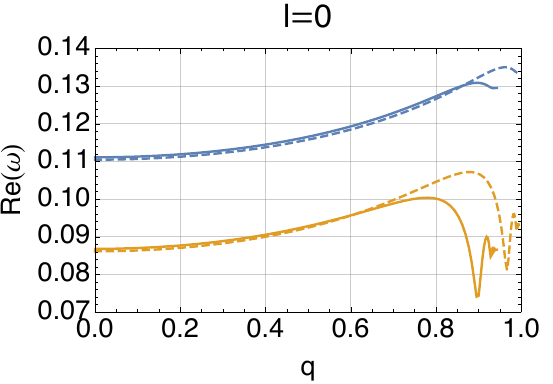}
		\includegraphics[height=4.6cm]{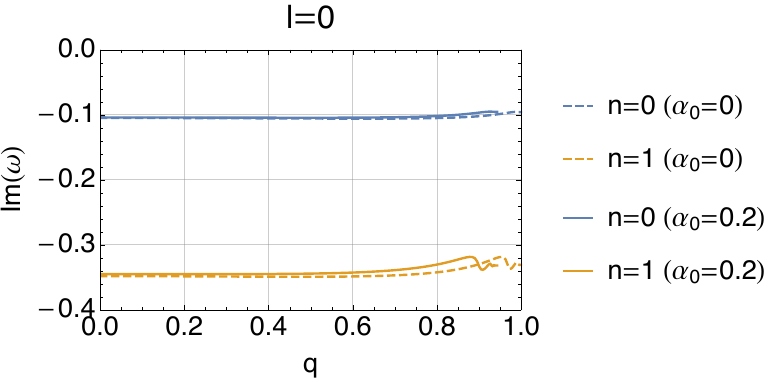}
		\caption{Real (left) and imaginary (right) parts of the scalar QNM frequencies as functions of $q$ for $l=0$. Solid and dashed curves correspond to Frolov BHs ($\alpha_0=0.2$) and RN BHs ($\alpha_0=0$), respectively. Blue and orange denote $n=0$ and $n=1$.}
		\label{Fro_FLvRNl0}
	}
\end{figure}

We begin with the charge dependence at $l=0$. Fixing $\alpha_0=0.2$ and scanning $q$, Fig.~\ref{Fro_FLvRNl0} compares Frolov QNFs (solid) with the RN limit $\alpha_0=0$ (dashed), while Table~\ref{Fro_FLvRNl0_Overtones} lists representative frequencies for $n=0$ and $n=1$. The fundamental mode varies smoothly. For $n=0$, $\mathrm{Re}(\omega)$ rises steadily with $q$ for both spacetimes, and the solid and dashed branches remain close until near extremality. The first overtone, by contrast, exhibits an overtone outburst \cite{Konoplya:2022pbc,Konoplya:2023kem}. The table shows that the Frolov $\mathrm{Re}(\omega)$ for $n=1$ grows from $\approx 0.087$ at $q=0$ to $\approx 0.100$ at $q=0.8$, then drops to $\approx 0.079$ at $q=0.89$, whereas the RN entry is still rising ($\approx 0.107$) at the same charge. Correspondingly, Fig.~\ref{Fro_FLvRNl0} shows the solid $n=1$ branch peaking near $q\approx 0.75$--$0.8$ and then falling sharply, while the dashed RN branch turns over only at larger $q$. A matching solid--dashed splitting appears in $\mathrm{Im}(\omega)$. The Frolov curves also terminate at smaller $q$ than their RN counterparts, so both the outburst threshold and the approach to extremality are shifted leftward relative to RN, consistent with \cite{Song:2024kkx}.


\begin{table}[!ht]
	\caption{Scalar QNM frequencies for the fundamental mode ($n=0$) and first overtone ($n=1$), with $l=0$ and $M=1$. Results are shown at selected values of $q$ for RN BHs ($\alpha_0=0$) and Frolov BHs ($\alpha_0=0.2$).}
	\centering
	\resizebox{16cm}{!}{
		\begin{tabular}{c|c|c|c|c}
			
			\hline
			&$\omega(q=0)$ & $\omega(q=0.6)$  &$\omega(q=0.8)$  &$\omega(q=0.89)$ \\ \hline
			$n=0(\alpha_0=0)$  &0.110455-0.104896i   &0.118455-0.105925i &0.126584-0.105047i &0.131970-0.102495i  \\ \hline
			$n=0(\alpha_0=0.2)$  &0.111077-0.104118i  &0.119286-0.104426i  &0.127300-0.101958i  &0.130833-0.097436i  \\ \hline
			$n=1(\alpha_0=0)$  &0.086117-0.348053i   &0.095603-0.348552i &0.104581-0.342206i  & 0.107062-0.331576i  \\ \hline
			$n=1(\alpha_0=0.2)$  & 0.086760-0.344856i   & 0.095651-0.342573i  &0.100146-0.330270i  & 0.078351-0.320588i  \\ \hline
	\end{tabular}}
	
	\label{Fro_FLvRNl0_Overtones}
\end{table}

\begin{figure}[H]
	\centering{
		\includegraphics[height=4.5cm]{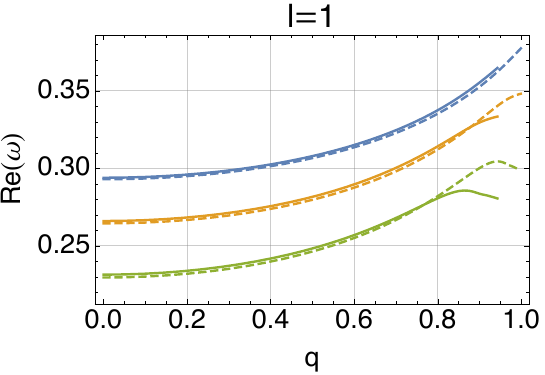}
		\includegraphics[height=4.6cm]{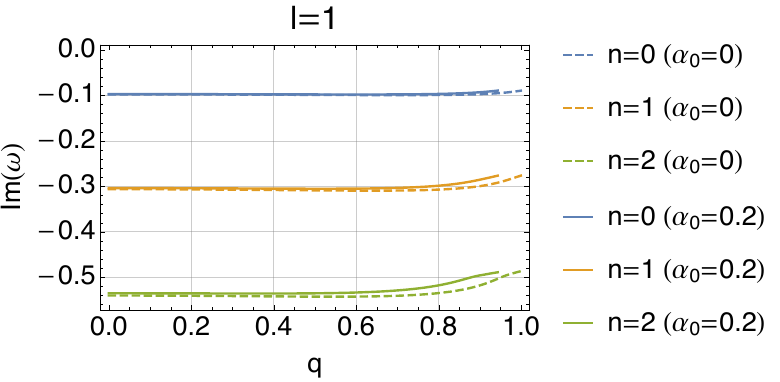}
		\caption{Real (left) and imaginary (right) parts of the scalar QNM frequencies as functions of $q$ for $l=1$. Solid and dashed curves correspond to Frolov BHs ($\alpha_0=0.2$) and RN BHs ($\alpha_0=0$), respectively. Blue, orange, and green denote $n=0,1,2$, respectively.}
		\label{Fro_FLvRNl1}
	}
\end{figure}

\begin{table}[!ht]
	\caption{Scalar QNM frequencies for the fundamental mode ($n=0$) and first overtone ($n=1$), with $l=0$ and $M=1$. Results are shown at selected values of $\alpha_0$ for Hayward BHs ($q=0$) and Frolov BHs ($q=0.6$).}
	\centering
	\resizebox{16cm}{!}{
		\begin{tabular}{c|c|c|c|c}
			
			\hline
			& $\omega(\alpha_0=0)$  &$\omega(\alpha_0=0.2)$  &$\omega(\alpha_0=0.4)$ & $\omega(\alpha_0=0.47)$ \\ \hline
			$n=0(q=0)$  &0.110455-0.104896i  &0.111077-0.104118i &0.112866-0.101402i  &0.113682- 0.099755i\\ \hline
			$n=0(q=0.6)$  &0.118455-0.105925i &0.119287-0.104427i  &0.121033-0.098595i  &0.120687-0.094671i \\ \hline
			$n=1(q=0)$  &0.086117-0.348053i &0.086760-0.344856i &0.087645-0.333839i  &0.087056-0.327244i \\ \hline
			$n=1(q=0.6)$  &0.095602-0.348552i & 0.095654-0.342574i  &0.087742-0.319939i  & 0.064691-0.311785i \\ \hline
	\end{tabular}}
	\label{Fro_FLvHwl0_Overtones}
\end{table}

We next ask how this charge-driven outburst changes when the angular momentum is raised from $l=0$ to $l=1$. Comparing Fig.~\ref{Fro_FLvRNl0} with Fig.~\ref{Fro_FLvRNl1}, one finds that raising $l$ lifts the whole $\mathrm{Re}(\omega)$ ladder---from roughly $0.08$--$0.13$ at $l=0$ to $0.22$--$0.30$ at $l=1$---as expected from the stronger centrifugal barrier, while every branch of $\mathrm{Im}(\omega)$ remains negative. More consequentially, the outburst onset shifts to a higher overtone index. At $l=0$ (Fig.~\ref{Fro_FLvRNl0}), as already noted, the $n=1$ curves peak and then drop sharply at large $q$. The Frolov branch turns over earlier than the RN branch, and this is accompanied by a matching splitting between the solid and dashed curves in $\mathrm{Im}(\omega)$. At $l=1$ (Fig.~\ref{Fro_FLvRNl1}), by contrast, the $n=0$ and $n=1$ branches remain nearly smooth and closely follow their RN counterparts until close to extremality. Only $n=2$ develops a clear peak-and-drop in $\mathrm{Re}(\omega)$ for the Frolov curves, again at smaller $q$ than in the RN limit. Angular momentum therefore does not eliminate the outburst, but postpones its onset from $n=1$ (at $l=0$) to $n\geq 2$ (at $l=1$), so that higher overtones must be tracked whenever $l>0$.


\begin{figure}[H]
	\centering{
		\includegraphics[height=4.6cm]{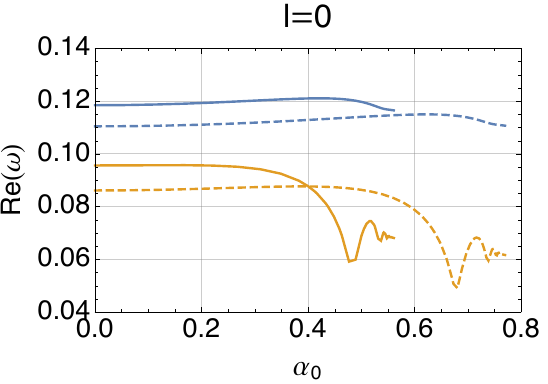}
		\includegraphics[height=4.6cm]{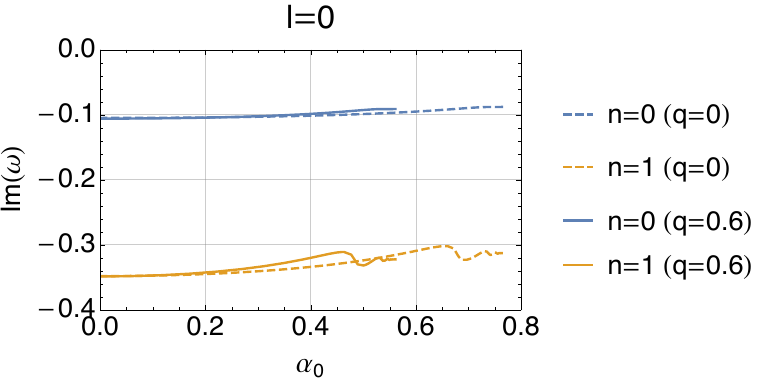}
		\caption{Real (left) and imaginary (right) parts of the scalar QNM frequencies as functions of $\alpha_0$ for $l=0$. Solid and dashed curves correspond to Frolov BHs ($q=0.6$) and Hayward BHs ($q=0$), respectively. Blue and orange denote $n=0$ and $n=1$.}
		\label{Fro_FLvHwl0}
}\end{figure}

\begin{figure}[H]
	\centering{
		\includegraphics[height=4.6cm]{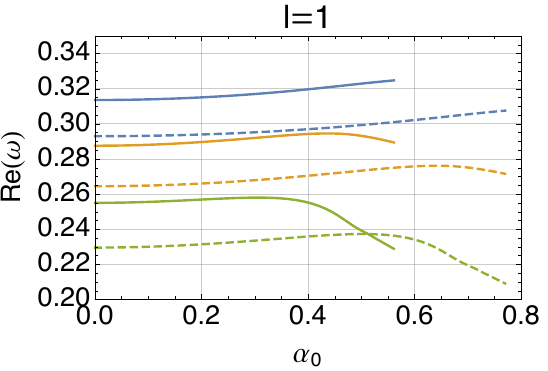}
		\includegraphics[height=4.6cm]{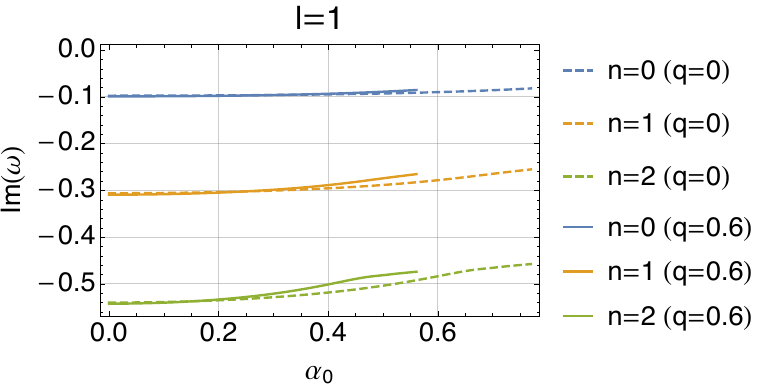}
		\caption{Real (left) and imaginary (right) parts of the scalar QNM frequencies as functions of $\alpha_0$ for $l=1$. Solid and dashed curves correspond to Frolov BHs ($q=0.6$) and Hayward BHs ($q=0$), respectively. Blue, orange, and green denote $n=0,1,2$, respectively.}
		\label{Fro_FLvHwl1}
}\end{figure}


An analogous outburst appears when the regularization parameter is varied instead of charge. With $q=0.6$ fixed, we scan $\alpha_0$ for $l=0$ and $l=1$; Table~\ref{Fro_FLvHwl0_Overtones} and Figs.~\ref{Fro_FLvHwl0} and \ref{Fro_FLvHwl1} compare Frolov BHs (solid) with the Hayward limit ($q=0$, dashed). Table~\ref{Fro_FLvHwl0_Overtones} and Fig.~\ref{Fro_FLvHwl0} ($l=0$) show that the fundamental mode again varies smoothly with $\alpha_0$, whereas the first overtone develops a clear outburst. For Frolov, $\mathrm{Re}(\omega)$ of $n=1$ drops from $\approx 0.096$ at $\alpha_0=0.2$ to $\approx 0.065$ at $\alpha_0=0.47$, while the corresponding Hayward entry remains near $0.087$ over the same range. In the figure, the solid $n=1$ branch turns over near $\alpha_0\approx 0.4$, whereas the dashed Hayward branch does so only near $\alpha_0\approx 0.6$. Thus the Frolov curves both terminate earlier and outburst at smaller $\alpha_0$, consistent with \cite{Gong:2023ghh}. Raising the angular momentum to $l=1$ (Fig.~\ref{Fro_FLvHwl1}) reproduces the same hierarchy found in the charge scan. The $n=0$ and $n=1$ modes stay nearly smooth and track the Hayward curves until close to the Frolov cutoff ($\alpha_0\approx 0.55$), while only $n=2$ exhibits a sharp peak-and-drop, again at smaller $\alpha_0$ than in the Hayward limit. Thus the $\alpha_0$-driven outburst mirrors the charge-driven one: Frolov turns over earlier than its classical counterpart, and angular momentum postpones the onset from $n=1$ (at $l=0$) to $n\geq 2$ (at $l=1$).

\begin{figure}[H]
	\centering{
		\includegraphics[height=6.2cm]{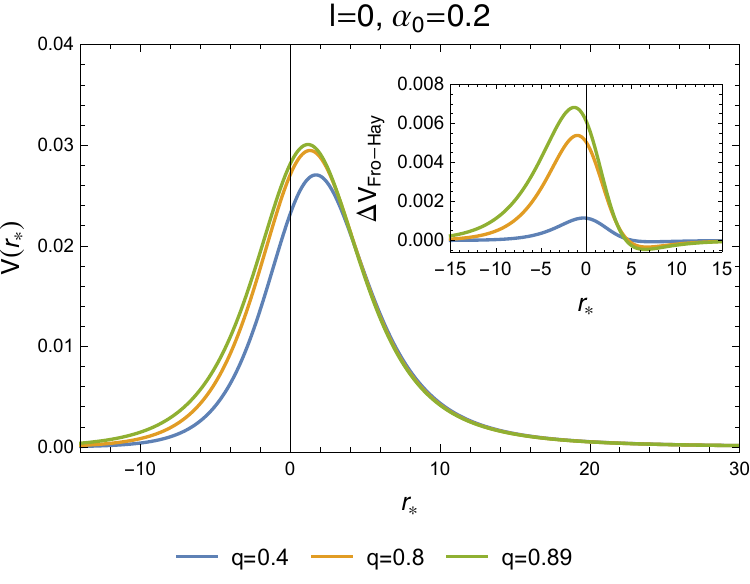}
		\includegraphics[height=6.1cm]{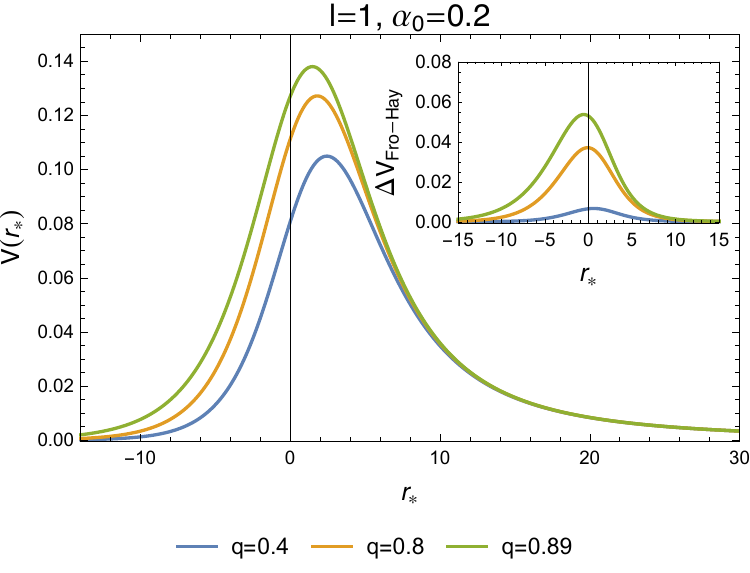}
		\caption{Effective potential $V(r_*)$ for scalar perturbations with $\alpha_0=0.2$ and $q=0.4,0.8,0.89$ (blue, orange, and green), for $l=0$ (left) and $l=1$ (right). The insets show $\Delta V_{\mathrm{Fro-Hay}}=V_{\mathrm{Fro}}-V_{\mathrm{Hay}}$, relative to the Hayward limit ($q=0$) at the same $\alpha_0$.}
		\label{Fro_Outburst}
	}
\end{figure}

\begin{figure}[H]
	\centering{
		\includegraphics[height=6.2cm]{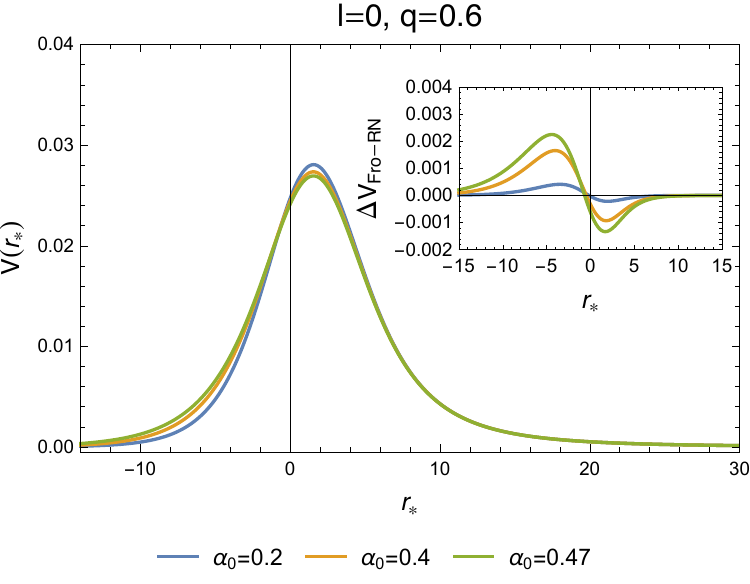}
		\includegraphics[height=6.1cm]{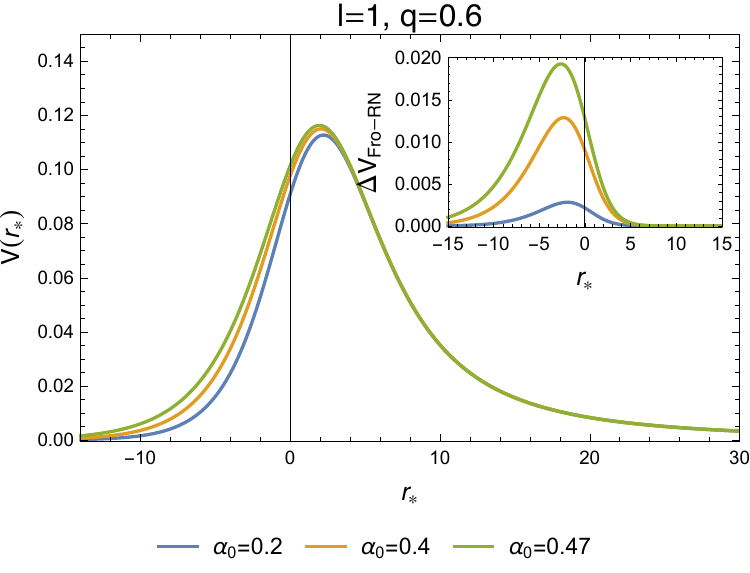}
		\caption{Effective potential $V(r_*)$ for scalar perturbations with $q=0.6$ and $\alpha_0=0.2,0.4,0.47$ (blue, orange, and green), for $l=0$ (left) and $l=1$ (right). The insets show $\Delta V_{\mathrm{Fro-RN}}=V_{\mathrm{Fro}}-V_{\mathrm{RN}}$, relative to the RN limit ($\alpha_0=0$) at the same $q$.}
		\label{Fro_Outburst2}
	}
\end{figure}


These spectral features find a natural explanation in the effective potential. Fig.~\ref{Fro_Outburst} shows $V(r_*)$ at fixed $\alpha_0=0.2$ for increasing $q$, with insets $\Delta V_{\mathrm{Fro-Hay}}$. As $q$ grows from $0.4$ to $0.89$, the barrier peak rises---mildly for $l=0$ (from $\approx 0.026$ to $\approx 0.029$) and much more strongly for $l=1$ (from $\approx 0.10$ to $\approx 0.13$)---while the peak drifts slightly toward the horizon. The insets localize the Frolov--Hayward difference near $r_*\lesssim 0$, with $\Delta V$ growing rapidly as $q$ approaches extremality. Fig.~\ref{Fro_Outburst2} repeats the comparison at fixed $q=0.6$ while scanning $\alpha_0$, now with insets $\Delta V_{\mathrm{Fro-RN}}$. Here the change in peak height is far milder and even $l$-dependent: for $l=0$ the peak decreases slightly as $\alpha_0$ increases, whereas for $l=1$ it rises only from $\approx 0.109$ to $\approx 0.112$. What both panels share is a near-horizon deformation relative to the chosen reference---Hayward in Fig.~\ref{Fro_Outburst} and RN in Fig.~\ref{Fro_Outburst2}---that strengthens with the scanned parameter. Charge therefore reshapes the barrier more strongly than $\alpha_0$, yet in either case the clearest deviation sits near the horizon---precisely where overtone wave functions are most sensitive. Overtone modes thus resolve horizon-scale differences of the Frolov geometry from both the classical RN limit and the regular Hayward limit, differences that remain largely invisible to the fundamental mode.

\section{Correspondence between grey-body factors and quasinormal modes}\label{secGBQNMs}

Using the QNFs computed in Sec.~\ref{secOvertone}, we next evaluate the grey-body factors via the QNM correspondence \cite{Konoplya:2024lir,Konoplya:2024vuj}. QNMs and grey-body factors describe the same radial problem but with different boundary conditions. QNMs satisfy Eq.~\eqref{BC-infinity}, whereas the grey-body factors obey
\begin{equation}
\begin{aligned}
\Psi &= e^{-i \Omega r_*} + R e^{i \Omega r_*}, \quad \quad \quad &r_*& \to +\infty\,, \\
\Psi &= T e^{-i \Omega r_*},  &r_*& \to -\infty\,.
\end{aligned}
\end{equation}
Here $R$ and $T$ are the reflection and transmission coefficients. The grey-body factor is then given by
\begin{equation}
    \Gamma_l(\Omega)=|T|^2 = 1-|R|^2\,.
\end{equation}


For spherically symmetric BHs, $\Gamma_l$ can be reconstructed from $\omega_0$ alone in the eikonal limit, but for small $l$ the first overtone $\omega_1$ must be retained \cite{Konoplya:2024lir}. Expanding the sixth-order WKB formula in $l^{-1}$ through $l^{-2}$, Konoplya and Zhidko obtained
\begin{equation}
     \Gamma_l(\Omega)=\frac{1}{1+e^{2\pi i \mathcal{K} }}\,,
\end{equation}
where
\begin{equation}
    - i \mathcal{K} = - \frac{\Omega^2 - \operatorname{Re}(\omega_0)^2}{4 \operatorname{Re}(\omega_0) \operatorname{Im}(\omega_0)} 
+ \Delta_1 + \Delta_2 + \Delta_f + \mathcal{O}(\ell^{-3})\,.
\label{eq:correspond}
\end{equation}

The first term on the right-hand side of Eq.~\eqref{eq:correspond} depends only on $\omega_0$; $\Delta_1$, $\Delta_2$, and $\Delta_f$ encode overtone corrections, with the remainder of $\mathcal{O}(l^{-3})$. Following \cite{Konoplya:2024lir}, the $\omega_0$-only truncation is the eikonal formula, the truncation that adds $\Delta_1$ is the first-order formula, and the expression retaining $\Delta_2$, and $\Delta_f$ is the amended formula. Because the underlying expansion is organized in powers of $l^{-1}$, we evaluate the amended formula at $l=1$ and examine how $q$ and $\alpha_0$ reshape $\Gamma_l$, as well as whether overtone corrections are required for accuracy in the Frolov background. For the amended formula, the explicit form of $i{\mathcal{K}}$ is
\begin{equation}
\begin{aligned}
i{\mathcal{K}} = &\frac{{{\Omega ^2} - {\mathop{\rm Re}\nolimits} {{\left( {{\omega _0}} \right)}^2}}}{{4{\mathop{\rm Re}\nolimits} \left( {{\omega _0}} \right){\mathop{\rm Im}\nolimits} \left( {{\omega _0}} \right)}}\left( {1 + \frac{{{{\left( {{\mathop{\rm Re}\nolimits} \left( {{\omega _0}} \right) - {\mathop{\rm Re}\nolimits} \left( {{\omega _1}} \right)} \right)}^2}}}{{32{\mathop{\rm Im}\nolimits} {{\left( {{\omega _0}} \right)}^2}}} - \frac{{3{\mathop{\rm Im}\nolimits} \left( {{\omega _0}} \right) - {\mathop{\rm Im}\nolimits} \left( {\omega 1} \right)}}{{24{\mathop{\rm Im}\nolimits} \left( {{\omega _0}} \right)}}} \right)\\
& - \frac{{{\mathop{\rm Re}\nolimits} \left( {{\omega _0}} \right) - {\mathop{\rm Re}\nolimits} \left( {{\omega _1}} \right)}}{{16{\mathop{\rm Im}\nolimits} \left( {{\omega _0}} \right)}} - \frac{{{{\left( {{\omega ^2} - {\mathop{\rm Re}\nolimits} {{\left( {{\omega _0}} \right)}^2}} \right)}^2}}}{{16{\mathop{\rm Re}\nolimits} {{\left( {{\omega _0}} \right)}^3}{\mathop{\rm Im}\nolimits} \left( {{\omega _0}} \right)}}\left( {1 + \frac{{{\mathop{\rm Re}\nolimits} \left( {{\omega _0}} \right)\left( {{\mathop{\rm Re}\nolimits} \left( {{\omega _0}} \right) - {\mathop{\rm Re}\nolimits} \left( {{\omega _1}} \right)} \right)}}{{4{\mathop{\rm Im}\nolimits} {{\left( {{\omega _0}} \right)}^2}}}} \right)\\
&  + \frac{{{{\left( {{\omega ^2} - {\mathop{\rm Re}\nolimits} {{\left( {{\omega _0}} \right)}^2}} \right)}^3}}}{{32{\mathop{\rm Re}\nolimits} {{\left( {{\omega _0}} \right)}^5}{\mathop{\rm Im}\nolimits} \left( {{\omega _0}} \right)}}\left( {1 + \frac{{{\mathop{\rm Re}\nolimits} \left( {{\omega _0}} \right)\left( {{\mathop{\rm Re}\nolimits} \left( {{\omega _0}} \right) - {\mathop{\rm Re}\nolimits} \left( {{\omega _1}} \right)} \right)}}{{4{\mathop{\rm Im}\nolimits}{{\left( {{\omega _0}} \right)}^2}}}} \right. \\
& \left. + {{\mathop{\rm Re}\nolimits} {{\left( {{\omega _0}} \right)}^2}\left( {\frac{{{{\left( {{\mathop{\rm Re}\nolimits} \left( {{\omega _0}} \right) - {\mathop{\rm Re}\nolimits} \left( {{\omega _1}} \right)} \right)}^2}}}{{16{\mathop{\rm Im}\nolimits} {{\left( {{\omega _0}} \right)}^4}}} - \frac{{3{\mathop{\rm Im}\nolimits} \left( {{\omega _0}} \right) - {\mathop{\rm Im}\nolimits} \left( {\omega 1} \right)}}{{12{\mathop{\rm Im}\nolimits} \left( {{\omega _0}} \right)}}} \right)} \right)+ \mathcal{O}(\ell^{-3})\,.
\end{aligned}
\end{equation}

\begin{figure}[H]
    \centering
    \begin{minipage}{0.45\textwidth}
        \centering
        \includegraphics[width=\linewidth]{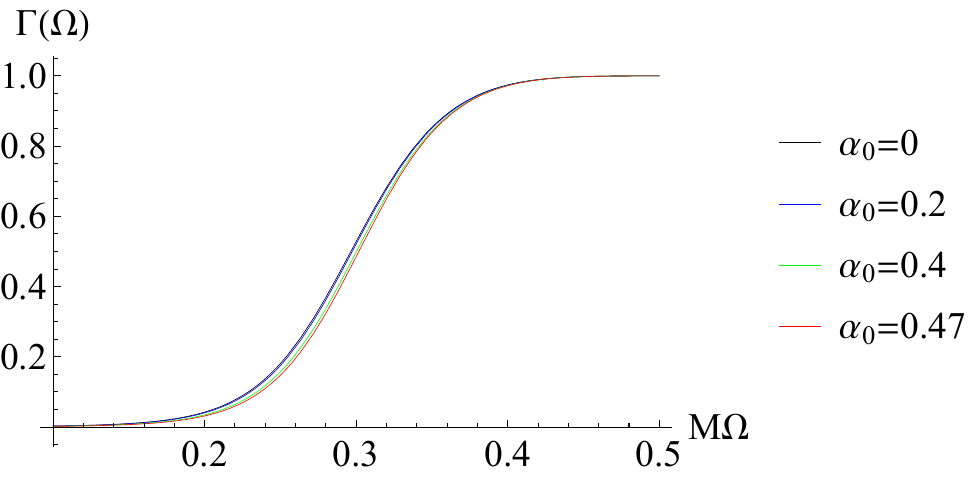}
    \end{minipage}
    \hfill
    \begin{minipage}{0.45\textwidth}
        \centering
        \includegraphics[width=\linewidth]{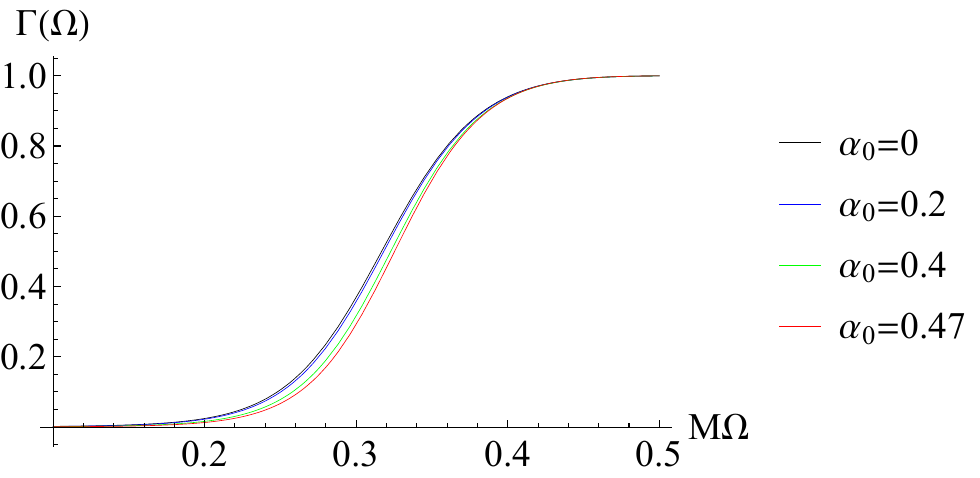}
    \end{minipage}
 \caption{Scalar grey-body factors $\Gamma_l(\Omega)$ as functions of $M\Omega$, obtained from the amended formula for $l=1$, with $q=0$ (left) and $q=0.6$ (right). Curves correspond to $\alpha_0=0,0.2,0.4,0.47$, as indicated in the legends.}
    \label{fig:GBl1q_amended formula}
\end{figure}
\begin{figure}[H]
    \centering
    \begin{minipage}{0.45\textwidth}
        \centering
        \includegraphics[width=\linewidth]{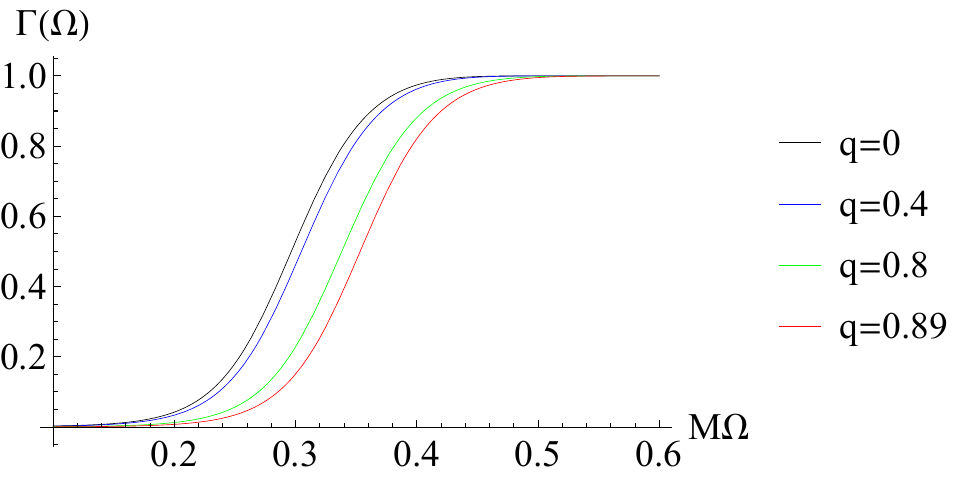}
    \end{minipage}
    \hfill
    \begin{minipage}{0.45\textwidth}
        \centering
        \includegraphics[width=\linewidth]{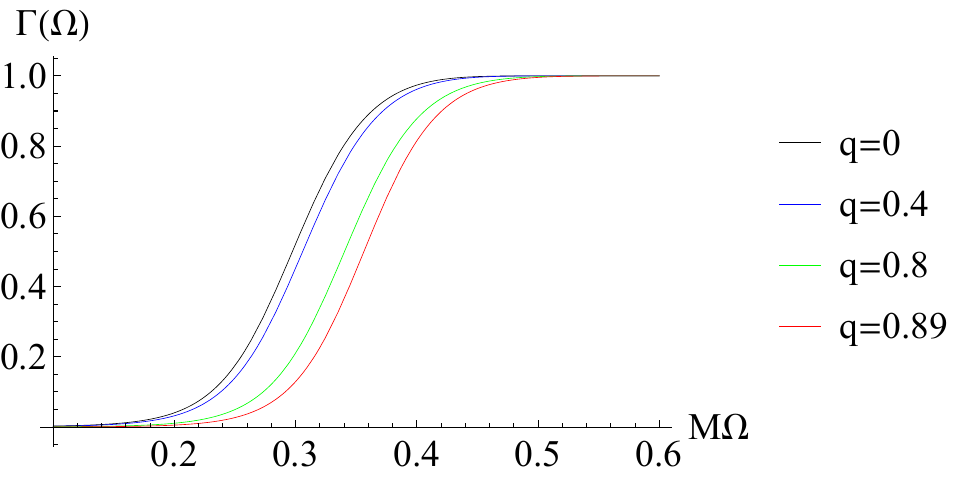}
    \end{minipage}
 \caption{Scalar grey-body factors $\Gamma_l(\Omega)$ as functions of $M\Omega$, obtained from the amended formula for $l=1$, with $\alpha_0=0$ (left) and $\alpha_0=0.2$ (right). Curves correspond to $q=0,0.4,0.8,0.89$, as indicated in the legends.}
    \label{fig:GBl1a_amended formula}
\end{figure}


Fig.~\ref{fig:GBl1q_amended formula} shows $\Gamma_l(\Omega)$ from the amended formula at $l=1$ while scanning $\alpha_0$ at fixed $q=0$ (left) and $q=0.6$ (right). Fig.~\ref{fig:GBl1a_amended formula} shows scans over $q$ at fixed $\alpha_0=0$ (left) and $\alpha_0=0.2$ (right). In every panel the spectrum has the usual sigmoid shape. $\Gamma_l$ stays near zero at low $M\Omega$, rises through an intermediate window, and saturates to unity at high frequency. Increasing either $\alpha_0$ or $q$ shifts the curves to the right, so that at fixed $\Omega$ the transmission is suppressed. The $q$-driven displacement in Fig.~\ref{fig:GBl1a_amended formula} is visibly larger than the $\alpha_0$-driven displacement in Fig.~\ref{fig:GBl1q_amended formula}. Raising $q$ from $0$ to $0.89$ moves the mid-rise of $\Gamma_l$ over a substantially wider frequency interval than raising $\alpha_0$ from $0$ to $0.47$. This hierarchy is consistent with the effective-potential analysis of Sec.~\ref{secOvertone}. For the $l=1$ barrier relevant here, increasing $q$ at fixed $\alpha_0=0.2$ raises the peak strongly (Fig.~\ref{Fro_Outburst}), whereas increasing $\alpha_0$ at fixed $q=0.6$ produces only a mild rise (Fig.~\ref{Fro_Outburst2}). A higher barrier impedes penetration and therefore lowers $\Gamma_l$ at fixed $\Omega$, with charge being the dominant factor.

\begin{figure}[H]
    \centering
    \begin{minipage}{0.45\textwidth}
        \centering
        \includegraphics[width=\linewidth]{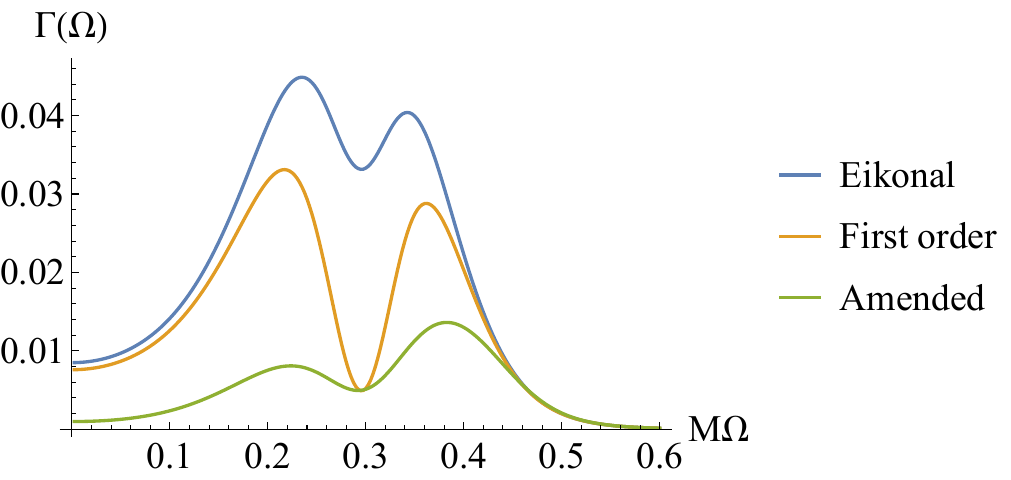}
    \end{minipage}
    \hfill
    \begin{minipage}{0.45\textwidth}
        \centering
        \includegraphics[width=\linewidth]{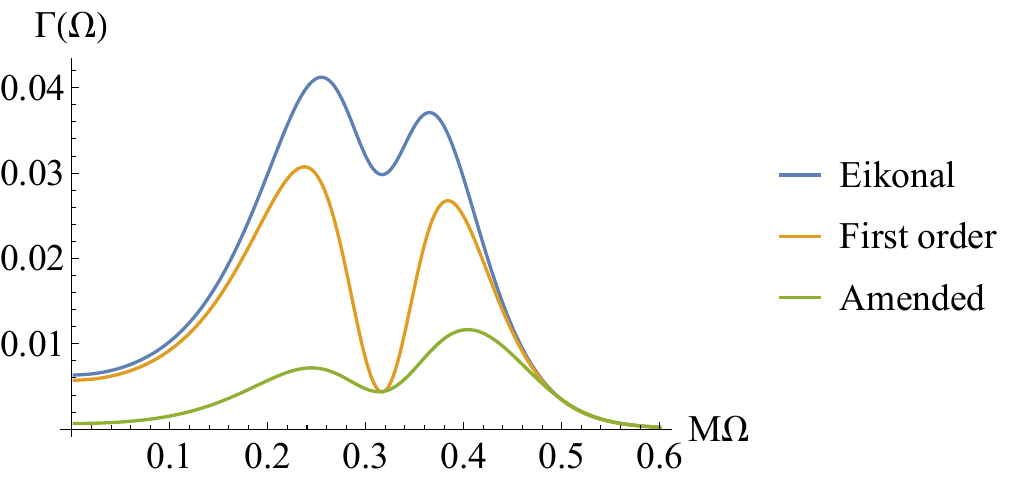}
    \end{minipage}
 \caption{Absolute errors $|\Gamma_l^{\mathrm{approx}}-\Gamma_l^{\mathrm{ref}}|$ of the eikonal (blue), first-order (orange), and amended (green) formulas relative to the reference grey-body factors, for $l=1$ and $\alpha_0=0.2$. The left and right panels correspond to $q=0$ and $q=0.6$, respectively.}
    \label{fig:GB_error}
\end{figure}

Fig.~\ref{fig:GB_error} compares the absolute deviations of the eikonal, first-order, and amended formulae from accurate grey-body factors at $l=1$, for $(\alpha_0,q)=(0.2,0)$ (left) and $(0.2,0.6)$ (right). In both panels, the eikonal curve (using only $\omega_0$) yields the largest error, with a double-peaked structure that reaches $\sim 0.04$--$0.045$ in the intermediate-frequency window. Adding the leading overtone correction reduces the error but preserves a similar double-peak pattern. The amended formula, which retains the full $\omega_1$ content through $\mathcal{O}(l^{-2})$, keeps the deviation $\lesssim 0.01$--$0.015$ across the plotted range and remains the smallest at both low and intermediate $M\Omega$. All three curves decay toward zero as $M\Omega\gtrsim 0.5$. Thus, for Frolov BHs at low $l$, overtone input is essential. The eikonal truncation is inadequate, whereas the amended formula provides a controlled and accurate reconstruction of $\Gamma_l$.

\section{Conclusions and discussions}\label{secconclusions}

We have computed the scalar overtone spectrum of Frolov black holes and the associated grey-body factors via the QNM correspondence. Within the allowed ranges of $q$ and $\alpha_0$, all modes satisfy $\omega_{\mathrm{Im}}<0$. This confirms stability under scalar perturbations and is consistent with the fundamental mode analysis in \cite{Song:2024kkx}.

Increasing either $q$ (at fixed $\alpha_0$) or $\alpha_0$ (at fixed $q$) can trigger overtone outbursts. These are characterized by a slow rise in $\omega_{\mathrm{Re}}$ followed by an abrupt drop beyond a critical parameter value. The same increase also shifts the outburst threshold to smaller parameter values. We tentatively attribute this shift to the fact that the increase drives the Frolov BH toward extremality earlier than in the RN ($\alpha_0=0$) or Hayward ($q=0$) limits. Angular momentum does not eliminate this phenomenon but postpones its onset. At $l=0$, the outburst already appears in $n=1$, whereas at $l=1$ one must track $n\geq 2$. Effective potential comparisons clarify the origin of these features. For $l=1$, increasing $q$ raises the barrier peak much more strongly than increasing $\alpha_0$ over the respective near-extremal scans, while in both scans the clearest deviations from the reference geometries are localized near the horizon. Here, the reference geometry is Hayward when $q$ is varied and RN when $\alpha_0$ is varied. Overtone modes are therefore sensitive probes of horizon-scale differences between the Frolov spacetime and both the classical RN limit and the regular Hayward limit. These differences remain largely invisible to the fundamental ringdown frequency.

Using the computed $\omega_0$ and $\omega_1$, we evaluated grey-body factors at $l=1$ via the QNM--grey-body correspondence for spherically symmetric backgrounds. Overtone corrections are essential at low angular momentum. The eikonal truncation (using only $\omega_0$) yields the largest error, whereas the amended formula, which retains the full $\omega_1$ content through $\mathcal{O}(l^{-2})$, keeps the deviation small across the relevant frequency window. At fixed $\Omega$, increasing either $q$ or $\alpha_0$ suppresses the transmission coefficient. Charge produces the stronger rightward shift of $\Gamma_l(\Omega)$, consistent with the larger barrier enhancement under the charge scan.

Future work could extend this analysis to electromagnetic and gravitational perturbations, higher multipoles, and rotating regular BH backgrounds. It could also explore whether overtone outbursts in Frolov-like geometries leave imprints on ringdown signals accessible to next-generation GW observatories.

\acknowledgments

This work is supported by the Natural Science Foundation of China under Grants No. 12375055.

\bibliographystyle{style1}
\bibliography{Ref}
\end{document}